\documentclass[letterpaper]{article} 
\usepackage[draft]{aa}  
\usepackage{times}  
\usepackage{helvet}  
\usepackage{courier}  
\usepackage[hyphens]{url}  
\usepackage{graphicx} 
\usepackage{natbib}  
\usepackage{caption} 
\usepackage{pdfpages}
\usepackage{algorithm}
\usepackage{algorithmic}
\usepackage{amssymb}
\usepackage{graphicx}
\usepackage{float}
\usepackage{subfig}
\usepackage{multirow}
\usepackage{enumitem}
\usepackage{amsmath}
\usepackage{lineno}
\usepackage{booktabs}
\usepackage{array}
\usepackage{textcomp}
\usepackage{newfloat}
\usepackage{listings}
\DeclareCaptionStyle{ruled}{labelfont=normalfont,labelsep=colon,strut=off} 
\floatstyle{ruled}
\newfloat{listing}{tb}{lst}{}
\floatname{listing}{Listing}

\title{Multimodal Difference Learning for Sequential Recommendation}
\author{
	Changhong Li\textsuperscript{\rm 1},
	Zhiqiang Guo\textsuperscript{\rm 2}\thanks{Corresponding author.},
	Guohui Li\textsuperscript{\rm 3},
	Li Zou\textsuperscript{\rm 1}
}
\affiliations{
	\textsuperscript{\rm 1}School of Computer Science and Technology, Huazhong University of Science and Technology, Wuhan, China\\
	\textsuperscript{\rm 2}Department of Computer Science and Technology, Tsinghua University, Beijing, China\\
	\textsuperscript{\rm 3}School of Software Engineering, Huazhong University of Science and Technology, Wuhan, China\\
	lichanghong.hust@gmail.com, zhiqiangguo@mail.tsinghua.edu.cn, \{guohuili, lizou\}@hust.edu.cn
}

\usepackage{bibentry}

\begin{document}

\maketitle

\begin{abstract}
Sequential recommendations have drawn significant attention in modeling the user's historical behaviors to predict the next item. With the booming development of multimodal data (e.g., image, text) on internet platforms, sequential recommendation also benefits from the incorporation of multimodal data. Most methods introduce modal features of items as side information and simply concatenates them to learn unified user interests. Nevertheless, these methods encounter the limitation in modeling multimodal differences. We argue that user interests and item relationships vary across different modalities.
To address this problem, we propose a novel \underline{M}ultimodal \underline{D}ifference Learning framework for \underline{S}equential \underline{Rec}ommendation, \textbf{MDSRec} for brevity. Specifically, we first explore the differences in item relationships by constructing modal-aware item relation graphs with behavior signal to enhance item representations. Then, to capture the differences in user interests across modalities, we design a interest-centralized attention mechanism to independently model user sequence representations in different modalities. Finally, we fuse the user embeddings from multiple modalities to achieve accurate item recommendation. Experimental results on five real-world datasets demonstrate the superiority of MDSRec over state-of-the-art baselines and the efficacy of multimodal difference learning.
\end{abstract}

\section{Introduction}
\label{sec:introduction}

Sequential recommender systems (SRSs) aim to uncover user preferences by analyzing the interaction sequences with temporal information.
Early Transform-based methods~\cite{sasrec,bert4rec} on SRSs generally take the ID information of items as sole data source. However, sparse interactions in real-world data hinder these methods from learning high-quality representations.
Recently, with the vigorous advancement of multimedia technology, a growing number of researchers have begun to explore incorporating multimodal data (e.g., images, texts, videos) of items into recommendation systems, achieving considerable achievements.

\begin{figure}[t]
	\centering
	\includegraphics[width=\linewidth]{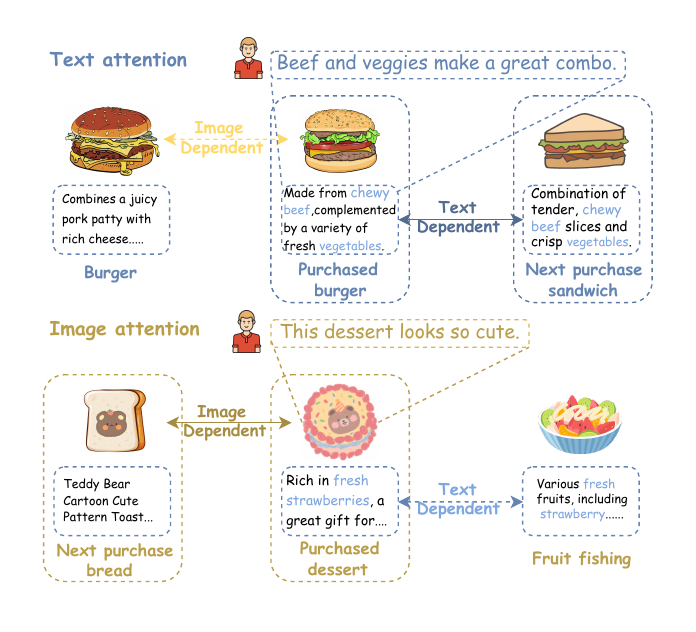}
	\caption{An illustration showing modality-related differences in user interests and item relationships.}
	\label{fig:prob}
\end{figure}

Many studies~\cite{vbpr,mmgcn} have demonstrated the significant value of integrating multimodal information in collaborative recommendation tasks. VBPR~\cite{vbpr} is the first work that introduces image information to enhance ID features of items. Subsequently, graph-based methods also benefit from multimodal integration. For instance, LATTICE~\cite{lattice} designs a modality-aware learning layer to explore latent semantic structure within modality features.
Although the exploration of multimodal data in sequential recommendation is still in its infancy, considerable progress has been made in its research.
For example, MMMLP~\cite{mmmlp} designs a multi-layer perceptron framework to simultaneously extract image, text, and item sequence information. MISSRec~\cite{missrec} tries to capture the sequence-level multimodal synergy and item-modality-interest relations for better sequence representation.

Despite the remarkable accomplishments, these methods still face challenges in modeling the differences between modalities.
(\textbf{i}) \textit{Differences in user interests across modalities}. 
Existing methods~\cite{odmt,mmsr} generally concatenate multimodal features of items within a sequence to represent the sequence, neglecting the differences in user interests across different modalities. As shown in Figure~\ref{fig:prob}, a user purchases a dessert and a burger, due to different interest attentions. For the burger, the user pays attention to the text description of its ingredients, i.e., beef and veggies, while the user thinks the visual appearance of the dessert looks cute. This provides evidence that a user's interests vary across different modalities. Simply combining the modal features of items in a user's sequence will poses challenges for modeling users’ unique interests across different modalities.
(\textbf{ii}) \textit{Differences in item relationships across modalities}.
Previous works~\cite{mmsrec,mmmlp} have almost entirely focused on modeling the modal features of items in sequence patterns, failing to capture the rich semantic relationships of items in multiple modalities. We argue that the item semantic relationships underlying different multimodal contents are beneficial for better item recommendation. Compared to other burgers in Figure~\ref{fig:prob}, the sandwich that is similar to the purchased burger in textual ingredients is more likely to be favored by the user. Likewise, the bread with an cute appearance like the dessert will become the next purchase of the user.
Therefore, the sequence pattern mining of modal features is limited and fails to model the rich and differentiated semantic relationships of items to enhance recommendation.

To address aforementioned issues, we propose a novel \underline{M}ultimodal-related \underline{D}ifference learning method for \underline{S}equential \underline{Rec}ommendation, which we term \textbf{MDSRec} for brevity.
Specifically, to explore the differences in item relationships across modalities, we construct item relationship graphs based on their modality features under each modality. Based on the learned relationship graphs, we perform graph convolutions to explicitly integrate high-order item affinities into item representations.
To mine the differences in user interests across modalities, we first cluster item modal features to obtain the modal-related interest centers. We then design an interest-centered attention mechanism to independently learn user preferences under each modality, in which we replace the original modal features of items with the learned item graph representations as input in the sequence.
Finally, we fuse the sequence embeddings from multiple modalities to obtain comprehensive user representations for item recommendation.
In summary, the main contributions of this paper are as follows:
\begin{itemize}[leftmargin=1.5em]
	\item We highlight the important of modeling the differences in user preferences and item relationships across modalities for multimodal sequential recommendation, which are help for discovering comprehensive user preferences.
	
	\item We propose a novel MDSRec framework for multimodal sequential recommendation, which mines modal-related item relationships and interest-centered user representations to learn modality differences in item relationships and user preferences, respectively.
	
	\item Extensive experiments on five real-world datasets demonstrate the superiority of our proposed model over state-of-the-art sequential recommendation baselines and validate the efficacy of modality difference learning in item relationships and user preferences.
	
\end{itemize}

\section{Related work}
\label{sec:relatedwork}

\begin{figure*}[t]
	\centering
	\begin{minipage}[t]{\textwidth}
		\centering
		\includegraphics[width=\textwidth]{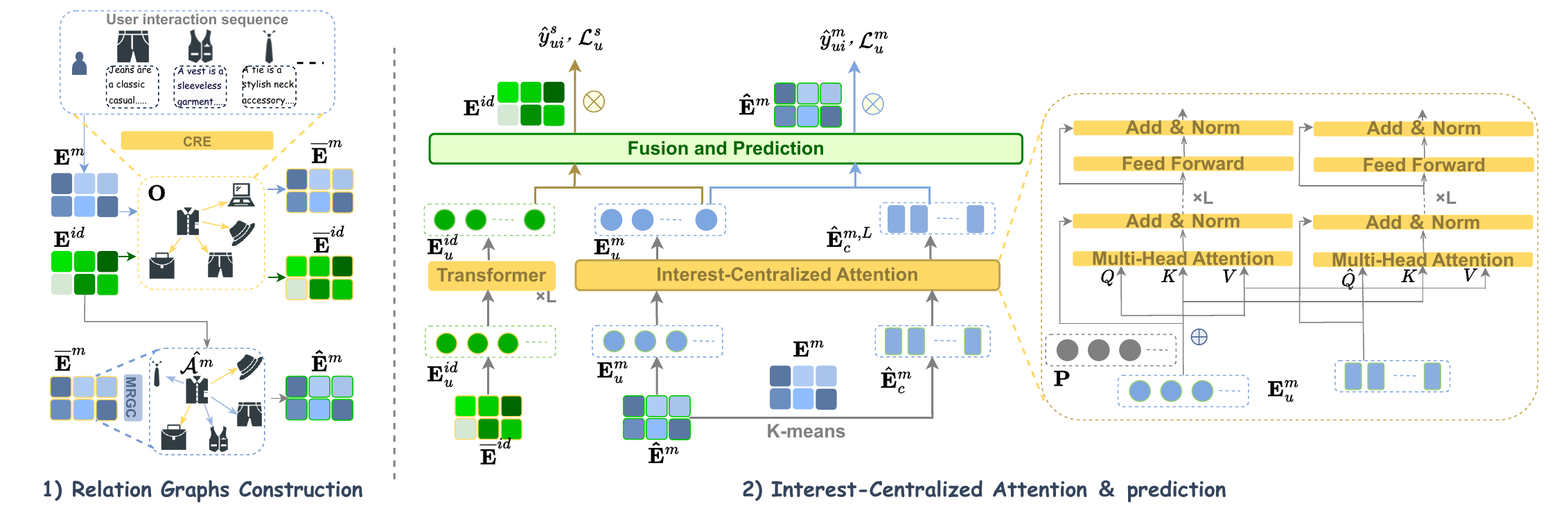}
	\end{minipage}
	\caption{The overall framework of MDSRec.}
	\label{fig:model}
\end{figure*}
\subsection{Sequential Recommendation}
Sequential recommendation aims to use the existing interaction sequences of users to predict the next most likely interacted item. Early sequential recommendations were mostly based on Markov chains~\cite{mark2,mark3} and pattern mining~\cite{pa1,pa2} that only obtain low-order simple dependency relationships. Subsequently, rapidly evolving neural networks have been introduced into recommendation systems. GRU4Rec~\cite{gru4rec} is a RNN-based model specifically designed for recommendation systems, using a variant of Gated Recurrent Unit. NARM~\cite{narm} proposes an RNN session with attention to extract long-term dependencies. The CNN-based Caser~\cite{caser} model can obtain collaborative information between items through convolutional filtering in two directions. Recently, the self attention mechanism of Transformer has been continuously applied in research of sequential recommendation. Compared to RNN, self-attention mechanism can capture behavior manner over longer distances. SASRec~\cite{sasrec} achieves excellent improvements by using self-attention to mine potential sequence behaviors of users. BERT4Rec~\cite{bert4rec} uses a bidirectional encoder to capture the preceding and following information of the sequence. However, since modality features are not introduced, their representation abilities are still limited by sparse interactions.

\subsection{Multi-modal Recommendation}
Multimodal recommendation systems have become the basic application on online platforms to provide personalized services to users. For traditional collaborative filtering recommendations, some methods~\cite{vbpr} directly use modality feature as side information to assist recommendations, while other methods~\cite{grcn,dualgnn,MGCN} utilize graph propagation techniques. Research on sequence recommendation is also abundant. FDSA~\cite{fdsa} utilizes attention mechanisms to capture a variety of heterogeneous product features. UniSRec~\cite{unisrec} offers a general method for sequence representation learning, using item text to derive more transferable representations. MMMLP~\cite{mmmlp} creates a multi-layer perceptron framework that concurrently extracts information from images, text, and item sequences. MissRec~\cite{missrec} introduces a novel framework for multimodal sequence recommendation, utilizing pre-training and transfer learning to effectively address the cold start problem and enable efficient domain adaptation. These approaches have achieved significant performance improvements and are highly representative and worth investigating. However, these methods generally concatenates multiple modal features to learn unified user interests, failing to exploring the differences between modalities, thereby achieving subpar performances.

\section{Methodology}
\label{sec:method}
\subsection{Notations and Problem Formulation}
We consider an implicit recommender system that consists of a user set $\mathcal{U}$ with $\mathcal{|U|}$ users, an item set $\mathcal{X}$ with $\mathcal{|X|}$ items and a modal set $\mathcal{M}=\{v,t\}$. The ID embeddings of items are denoted as $\mathbf{E}^{id}=\{\mathbf{e}_1^{id}, \dots , \mathbf{e}_i^{id}, \dots ,  \mathbf{e}^{id}_\mathcal{|X|}\} \in \mathbb{R}^{\mathcal{|X|} \times {d}}$. The modal features of items are represented as $\mathbf{E}^{m}=\{\mathbf{e}_1^m, \dots , \mathbf{e}_i^m, \dots ,  \mathbf{e}^m_\mathcal{|X|}\} \in \mathbb{R}^{\mathcal{|X|} \times {d_m}}$, $m \in \mathcal{M}$. Each user is represented by their own interaction history sequence $\mathcal{S}^{u}=\{{x}_1, \dots ,{x}_i, \dots , {x}_t| {x}_i \in \mathcal{X}\}$, $u \in \mathcal{U}$, where ${t}$ represents the set sequence length. Based on a given user interaction sequence $\mathcal{S}^{u}$. The core goal of sequential recommendation is to predict the next item that the user is most likely to interact with.

\subsubsection{Overview} Figure~\ref{fig:model} illustrates the overall framework of MDSRec, which contains three main parts: 
1) an item relation graph construction (RGC) module that constructs multiple item relation graphs through co-occurrence information and modal features to enhance item representations.
2) an interest-centralized attention (ICA) module that dependently models user interests across modalities by jointing Transformer architecture and centralized attention mechanism.
3) a fusion and prediction module that fuses the user preferences across modalities to achieve item recommendation.
\subsection{Item Relation Graph Construction}
To extract the differences in item relationships across modalities, we construct the item relation graphs based on the modal features of items. These neighbors that are semantically similar to an item may become potential interactions for a user who interact with the item. Besides, we incorporate sequential co-occurrence information into the item relation graph construction process to strengthen the connection between modalities and behavioral signals.

\subsubsection{\textbf{Co-occurrence Relation Extraction(CRE)}}
Sequential co-occurrence relation implies behavior-related item collaboration information. We aim to inject the behavioral signals into modal features to enhance the robustness of item relationship modeling. Therefore, we first extract the item co-occurrence relation to capture behavioral signals.
Specifically, for two items ${x_i}$ and ${x_j}$ in a sequence, we believe that the closer their relative distance, the stronger their relationship tends to be. Thus, we calculate the behavioral affinity score $\mathrm{O}^u_{ij}$ between items ${x_i}$ and ${x_j}$ for user $u$ as,
\begin{equation}
	\mathrm{O}^u_{ij}= 
	\begin{cases}
		\frac{1}{\mathrm{D}_{ij}}, & \mathrm{if} ~~~ x_i \in \mathcal{S}^{u}, x_j \in \mathcal{S}^{u},\\
		0, & \mathrm{otherwise},
	\end{cases}
\end{equation}
where $\mathrm{D}_{ij}$ represents the positional distance between items ${x_i}$ and ${x_j}$ in ser $u$'s sequence. Then, we sum behavioral affinity scores between items ${x_i}$ and ${x_j}$ in all users' sequence to obtain the their final co-occurrence score $\mathrm{O}_{ij}$,
\begin{equation}
	\mathrm{O}_{ij} =\sum_{u\in\mathcal{U}}{\mathrm{O}^u_{ij}}.
\end{equation}
By performing similar calculations for all pairs of items, we obtain item co-occurrence relation matrix $\mathbf{O} \in \mathbb{R}^{\mathcal{|X|} \times \mathcal{|X|}}$. 

Thereafter, we inject behavioral signals into item modal representations,
\begin{equation}
	\overline{\mathbf{E}}^n = \mathbf{\mu}^n \cdot  \mathbf{O}{\mathbf{E}}^n + \mathbf{E}^n,
\end{equation}
where we set $n \in \mathcal{M}\cup\{id\}$. $\mathbf{\mu}^{n}$ is a adjust parameter to control the degree of injection of behavioral signals. $\overline{\mathbf{E}}^{n}$ is a feature matrix with co-occurrence information. Here, $\overline{\mathbf{E}}^{m}$ is the modal feature matrix of items that is used to construct subsequent item relation graph, where $m \in \mathcal{M}$.  $\overline{\mathbf{E}}^{id}$ is ID embedding matrix of items for subsequent representation learning.


\subsubsection{\textbf{Modal-aware Relation Graph Construction(MRGC)}}
In order to capture the semantic differences between modalities, we attempt to construct item relation graphs in different modalities. 
Specifically, we adopt the cosine similarity to calculate the semantic affinity of two items ${x_i}$ and ${x_j}$,
\begin{equation}
	{\mathcal{A}}^m_{ij} = \frac{(\overline{\mathbf{\boldmath{e}}}^m_i)^{\top}(\overline{\mathbf{\boldmath{e}}}^m_j)}
	{||\overline{\mathbf{\boldmath{e}}}^m_i||||\overline{\mathbf{\boldmath{e}}}^m_j||},
\end{equation}
	where ${\mathcal{A}}^m_{ij}$ represents the semantic affinity score between items ${i}$ and ${j}$ in modality $m$. $\overline{\mathbf{e}}^m_i$ and $\overline{\mathbf{e}}^m_j$ are modal features of items ${i}$ and ${j}$ extracted from the matrix $\overline{\mathbf{E}}^{m}$. For item ${x_i}$, its semantic similarity with all items in modality ${m}$ can be expressed as ${\mathcal{A}}^m_{i*} = [{\mathcal{A}}^m_{i1};\dots ; {\mathcal{A}}^m_{ij}; \dots ; {\mathcal{A}}^m_{i\mathcal{|X|}}]$. Then, we select top-$H$ items with highest scores as the neighboring items of item ${x_i}$, and set their affinity score to $1$,
\begin{equation}
	\mathcal{\hat{A}}^m_{ij}= 
	\begin{cases}
		1, & {\mathcal{A}}^m_{ij}\in \mathrm{top-}{H}({\mathcal{A}}^m_{i*}),\\
		0, & \mathrm{otherwise},
	\end{cases}
\end{equation}
where $H$ is hyperparameter. By performing the above procedure for each item, we can construct the semantic affinity graph $\mathcal{\hat{A}}^m \in \mathbb{R}^{\mathcal{|X|} \times \mathcal{|X|}}$ as item relation graph for modality $m$. Then, we adopt one-layer light graph convolutional network to obtain the semantic features $\mathbf{\hat{E}^m}$ of items,
\begin{equation}
	\mathbf{\hat{E}^m} = {\mathcal{\hat{A}}^m}\mathbf{E}^{id}.
\end{equation}
Here, we transfer the semantic signals of modal features to the ID embeddings.

\subsection{Interest-Centralized Attention}
Following recent methods~\cite{sasrec}, we utilize Transformer~\cite{transformer} to learn accurate and reliable sequence representations. To further model the differences in user interests across modalities, we introduce interest-centralized attention mechanism to extract user preferences within a modality.
\subsubsection{\textbf{User Sequence Representation Learning}}
Transformer~\cite{transformer} is highly suitable for the problem scenario of sequence recommendation, and we use it to capture long-distance dependencies in sequence embeddings. Firstly, taking the learned ID embeddings $\overline{\mathbf{E}}^{id}$ as input, we introduce positional information for the sequence,
\begin{equation}
	\begin{aligned}
		\mathbf{E}^{id}_{u}&=\overline{\mathbf{E}}^{id}[\mathcal{S}^u]+\mathbf{P} \\
		&=[\overline{\mathbf{e}}_1^{id}+\mathbf{p}_1,\ldots, \overline{\mathbf{e}}_i^{id}+\mathbf{p}_i, \ldots, \overline{\mathbf{e}}_{t}^{id}+\mathbf{p}_{t}],
	\end{aligned}
\end{equation}
where $\overline{\mathbf{E}}^{id}[\mathcal{S}^u]$ represents the extraction of item ID embeddings in user $u$'s sequence. $\mathbf{P}=\{\mathbf{p}_1,\dots , \mathbf{p}_i, \dots , \mathbf{p}_{t}\}$ is position vector. Then, after applying operations like masking, $\mathbf{E}^{id}_{u}$ can be fed into the transform for learning.
\begin{equation}
	\mathbf{E}^{id}_{u}=\mathbf{Trm}^{L}(\mathbf{E}^{id}_{u}),
\end{equation}
where $\mathbf{Trm}(\cdot)$ is a Transformer block and $L$ is the block number. $\mathbf{E}^{id}_{u}$ is the user sequence embeddings. Similarly, we obtain the user modal-related preference representation by treating the modal features $\hat{\mathbf{E}}^{m}$ of items as input,


\begin{equation}
	\begin{aligned}
		\mathbf{E}^{m}_{u}&=\hat{\mathbf{E}}^{m}[\mathcal{S}^u]+\mathbf{P},\\
		\mathbf{E}^{m}_{u}&=\mathbf{Trm}^{L}(\mathbf{E}^{m}_{u}) 
	\end{aligned}
\end{equation}

\subsubsection{\textbf{Centralized Attention}}
Existing works~\cite{mmsr,mmmlp} lack in-depth exploration of user interests across modalities. To uncover more accurate and reliable user interests, we design a centralized attention module. Specifically, we first obtain feature centers for each modality through k-means~\cite{k1,k2} clustering,
\begin{equation}
	\mathbf{C}^m = \mathrm{k-means}(\mathbf{E}^m),
\end{equation}
where $\mathbf{C}^m \in \mathbb{R}^{k \times \mathcal{|X|}}$ represents the relationship between all items and $k$ cluster centers towards modality $m$. Then, we compute the center features $\hat{\mathbf{E}}^m_c$ by,
\begin{equation}
	\hat{\mathbf{E}}^m_c = \mathbf{C}^m\mathbf{\hat{E}}^m.
\end{equation}
Further, $\hat{\mathbf{E}}^m_c$ is input into our designed centralized attention module to learn key user interests during the representation learning process. For modality $m$, the process of updating center feature is as follows,
\begin{equation}
	\mathbf{a}^l_h = \textsc{Softmax}\left(\frac{(\hat{\mathbf{E}}_c^{m,l-1}\mathbf{Q}^l_h)(\mathbf{E}_u^{m, l-1}\mathbf{K}^l_h)^T}{\sqrt{d}}\right),
\end{equation}
\begin{equation}
	\mathbf{head}^l_h =\mathbf{a}^l_h(\mathbf{E}_u^{m, l-1}\mathbf{V}^l_h),
\end{equation}
\begin{equation}
	\mathbf{g}^l =[\mathbf{head}_1^l; \mathbf{head}_2^l; \dots; \mathbf{head}_{|{h}|}^l]\mathbf{U}^l,
\end{equation}
\begin{equation}
	\hat{\mathbf{E}}_c^{m,l} = \sigma((\mathbf{g}^l\mathbf{W}^l_1+\mathbf{b}^l_1)\mathbf{W}^l_2+\mathbf{b}^l_2),
\end{equation}
where $\hat{\mathbf{E}}_c^{m,l}$ represents the center feature of $l$-th layer, and $\hat{\mathbf{E}}_c^{m,0}=\hat{\mathbf{E}}_c^{m}$.
$\mathbf{Q}^l_h$, $\mathbf{K}^l_h$ and $\mathbf{V}^l_h \in \mathbb{R}^{{d_m} \times {d_m}}$ share the multi-head attention weight of Transformer in sequence representation learning to generate the query, key and value vectors.
$\mathbf{E}_u^{m, l-1}$ represents the output of the Transformer at $(l-1)$-th layer. $\mathbf{a}^h_{l}$ is the generated attention score of the $h$-th attention head. $h$ is the number of heads. After $L$ layers of centralized attention learning, the final center representations are updated as,
\begin{equation}
	\mathbf{E}^m_c = \hat{\mathbf{E}}_c^{m,L}.
\end{equation}
Here, $\mathbf{E}^m_c$ is able to capture the user main interests, uncovering differences in user interests across modalities.

\subsection{Fusion and Prediction}
\subsubsection{Representation Fusion}
Considering that the last item in the sequence often has a high correlation with predicting the next item, we fuse the sequence representations from multiple modalities to explore a more comprehensive understanding of user interests,
\begin{equation}
	\mathbf{e}_u^s = \sum_{m\in\mathcal{M}}\rho_m \cdot \mathbf{e}^m_{u,t}+{\mathbf{e}}^{id}_{u,t},
\end{equation}
where $\mathbf{e}^m_{u,t}$ and ${\mathbf{e}}^{id}_{u,t}$ are the last ($t$-th) item representations in user $u$'s sequence. $\rho_m$ is a hyperparameter used to adjust the integration of modal features. We set $\sum_{m\in\mathcal{M}} \rho_m = 1$.

To capture accurate user interest differences, we further integrate generated center features into the modal embeddings,
\begin{equation}
	\widetilde{\mathbf{E}}_u^m = \mathbf{E}_u^m + \mathbf{\Gamma} \mathbf{E}^m_{c},
\end{equation}
where $\widetilde{\mathbf{E}}_u^m$ is the modal feature matrix with user center interests. $\mathbf{\Gamma}$ is the relation matrix between modal embeddings of items and center representations. We employ the Gumbel-Softmax~\cite{gumbel} function to implement its calculation, 
\begin{equation}
	\mathbf{\boldsymbol{\gamma}}^u =\textsc{Softmax}\left(\frac{\log\boldsymbol{\delta}-\log(1-\boldsymbol{\delta})+\mathbf{e}_{u}^m \mathbf{E}_{c}^{m \top}}{\tau}\right),
\end{equation}
where $\mathbf{\boldsymbol{\gamma}}^u \in \mathbb{R}^k$ is the $u$-th relation vector in $\mathbf{\Gamma}$. $\boldsymbol{\delta} \in \mathbb{R}^k$ is
a noise vector, where each value $\delta_a \sim \text{Uniform}(0,1)$, and $\tau$ is a temperature weight. Similarly, we choose the last item modal feature $\widetilde{\mathbf{e}}_{u,t}^m$ for next item prediction.

\subsubsection{Prediction and Optimization}
After obtaining the user sequence representation, we use the user representation and ID embeddings of items to calculate the prediction score $\hat{y}_{ui}^s$ of user $u$ and item $x_i$,
\begin{equation}
	\hat{y}_{ui}^s =\mathbf{e}_u^s(\mathbf{e}_{i}^{id})^{\top},
\end{equation}
where $\mathbf{e}_{i}^{id}$ and is the ID embedding of item $x_i$. 

However, predicting solely based on the fused sequence embeddings from multiple modalities lacks independent modeling of user interest variations. 
Therefore, we achieve the independent prediction in each modality by utilizing centralized modal features $\widetilde{{\mathbf{e}}}_{u}^m$ of user $u$,
\begin{equation}
	\hat{y}_{ui}^m =\widetilde{{\mathbf{e}}}_{u}^m(\hat{\mathbf{e}}^m_i)^{\top},
\end{equation}
where $\hat{y}_{ui}^m$ is the predicted score for user $u$ and item $x_i$ in modality $m$. This approach allows for a more accurate extraction of user preferences in each modality and further uncover the differences in user interests across modalities.

Thereafter, the final prediction score $\hat{y}_{ui}$ of user $u$ and item $x_i$ is calculated as,
\begin{equation}
	\hat{y}_{ui} =\hat{y}_{ui}^s + \sum_{m\in\mathcal{M}}{\rho_m \cdot \hat{y}_{ui}^m}.
\end{equation}

Subsequently, following other sequential recommendations~\cite{odmt,missrec}, we use cross entropy loss~\cite{cross1,cross2} as the recommendation loss, which can minimize the negative logarithmic likelihood of the base truth value for correctly recommending the next item. Based on the prediction results mentioned earlier, we optimize our model via the cross entropy loss,
\begin{equation}
	\mathcal{L} = \frac{1}{|\mathcal{U}|} \sum_{u\in\mathcal{U}} \mathcal{L}_u^s + \sum_{m\in\mathcal{M}}{\rho_m \cdot \mathcal{L}_u^m},
\end{equation}
where the loss $\mathcal{L}_u^m$ is implemented by,
\begin{equation}
	\mathcal{L}_u^m = -\sum_{x_i\in\mathcal{X}}y_{ui}\log(\hat{y}^m_{ui}),
\end{equation}
where $y_{ui}$ is the ground-truth binary interaction value. The loss $\mathcal{L}_u^s$ can be implemented in similar manner. Notice we calculate the final prediction loss by fusing the cross entropy losses about all prediction channel, which is beneficial for reliability modeling of user interest in modalities.

\section{Experiment}
\label{sec:experiment}
\begin{table}[t]
  \centering
   \setlength{\tabcolsep}{1.5mm}{
    \begin{tabular}{lrrrrc}
    \toprule
    \textbf{Dataset} & \textbf{\#Users} & \textbf{\#Items} & \textbf{\#Inters} & \textbf{Avg.n} & \multicolumn{1}{c}{\textbf{Sparsity}}\\
    \midrule
    \textbf{Scientific}  & 8,443 & 4,386 & 50,985   & 6.039  & 0.9986\\
    \textbf{Pantry} & 13,102 & 4,899 & 113,861  & 8.691  & 0.9982\\
    \textbf{Baby}    & 19,446  & 7,051  & 141,347  & 7.269  & 0.9990\\
    \textbf{Sports}  & 35,599 & 18,358 & 260,739   & 7.325  & 0.9996\\
    \textbf{Clothing} & 39,388 & 23,034 & 239,290  & 6.075  & 0.9997\\
    \bottomrule
    \end{tabular}}
  \caption{Statistics of five evaluation datasets.}
  \label{tab:dataset}
\end{table}
\setlength{\tabcolsep}{1.2mm}
\begin{table*}[t]
	\centering
	\small
	  \setlength{\tabcolsep}{0.9mm}{
	\begin{tabular}{llccclcccccccr}
		\toprule
		\multicolumn{1}{l}{\multirow{2}{*}{Datasets}}   & \multirow{2}{*}{Metric} & \multicolumn{3}{c}{ID-based SR}                                    &  & \multicolumn{7}{c}{Modality-based SR}                                               &  \multirow{2}{*}{improv.} \\ \cmidrule(l){3-5} \cmidrule(l){7-13} 
		\multicolumn{1}{c}{}                            &                         & \raisebox{-0.5ex}[0pt]{BERT4Rec}             & \raisebox{-0.5ex}[0pt]{GRU4Rec}              & \raisebox{-0.5ex}[0pt]{SASRec}             &   & \raisebox{-0.5ex}[0pt]{GRU4RecF}             & \raisebox{-0.5ex}[0pt]{SASRecF}              & \raisebox{-0.5ex}[0pt]{FDSA}                 & \raisebox{-0.5ex}[0pt]{UniSRec}               & \raisebox{-0.5ex}[0pt]{MMMLP}                & \raisebox{-0.5ex}[0pt]{MissRec}              & \raisebox{-0.5ex}[0pt]{MDSRec}              &                                   \\ 
		\midrule
		\multicolumn{1}{l}{\multirow{4}{*}{Scientific}} & R@10                    & 0.0454          & 0.0666        & 0.0842       &                      & 0.0964          & 0.1145         & 0.0892       & 0.1311           & 0.1019             & \textbf{0.1360}              & \underline{0.1359}             &  -0.07\%                                 \\
 		\multicolumn{1}{l}{}                            & N@10                    & 0.0232          & 0.0372        & 0.0466       &                      & 0.0615          & 0.0597         & 0.0573       & 0.0658                        & 0.0648             & \underline{0.0753}              & \textbf{0.0774}             &       2.79\%                           \\
		\multicolumn{1}{l}{}                            & R@20                    & 0.0682         & 0.9590        & 0.1151       &                      & 0.1233          & 0.1496         & 0.1160       & 0.1745                          & 0.1346             & \underline{0.1748}              & \textbf{0.1761}             &    0.74\%                                 \\
		\multicolumn{1}{l}{}                            & N@20                    & 0.0288          & 0.0445        & 0.0542       &                      & 0.0681          & 0.0683         & 0.0639       & 0.0766                        & 0.0727             & \underline{0.0839}              & \textbf{0.0873}             &   4.05\%                                  \\ \midrule
		\multirow{4}{*}{Pantry}                         & R@10                    & 0.0356          & 0.0395        & 0.0434       &                      & 0.0481          & 0.0601         & 0.0434       & 0.0763                            & 0.0521             & \underline{0.0779}              & \textbf{0.0822}             &  5.52\%                                   \\
		& N@10                    & 0.0168          & 0.0190        & 0.0211       &                      & 0.0240          & 0.0256         & 0.0215       & 0.0360                          & 0.0255             & \underline{0.0365}             & \textbf{0.0391}             &     7.12\%                                \\
		& R@20                    & 0.0582          & 0.0687        & 0.0723       &                      & 0.0776          & 0.0885         & 0.0717       & 0.1149                             & 0.0824             & \underline{0.1158}              &\textbf{0.1201}            &   3.71\%                                  \\
		& N@20                    & 0.0223          & 0.0261        &  0.0276     &                      &0.0313           & 0.0326         & 0.0285       & 0.0454                            & 0.0329             & \underline{0.0458}              & \textbf{0.0484}             &                5.68\%                    \\ \midrule		
		\multirow{4}{*}{Baby}                           & R@10                    & 0.0254               & 0.0445               & 0.0447               &  & 0.0446               & 0.0474               & 0.0476               & \underline{0.0534}                & 0.0453               & 0.0519               &    \textbf{0.0584}               &              9.36\%                      \\
		& N@10                    & 0.0121               & 0.0218               & 0.0216               &  & 0.0212               & 0.0202               & 0.0228               & 0.0240               & 0.0220               & \underline{0.0247}               & \textbf{0.0285}               &        15.38\%                             \\
		& R@20                    & 0.0428               & 0.0699               & 0.0717               &  & 0.0723               & 0.0754               & 0.0750               & \underline{0.0848}                & 0.0718               & 0.0787               & \textbf{0.0905}               &   6.72\%                                  \\
		& N@20                    & 0.0164               & 0.0279               & 0.0282               &  & 0.0279               & 0.0271               & 0.0295               & \underline{0.0314}                & 0.0285               & 0.0313               & \textbf{0.0364}               &      15.92\%                               \\ \midrule
		\multirow{4}{*}{Sports}                         & R@10                    & 0.0230               & 0.0398               & 0.0423               &  & 0.0435               & 0.0542               & 0.0433               & \underline{0.0597}                & 0.0416              & 0.0591               & \textbf{0.0664}               &      11.22\%                               \\
		& N@10                    & 0.0114               & 0.0198               & 0.0208               &  & 0.0210               & 0.0236              & 0.0214               & 0.0262                & 0.0203               & \underline{0.0281}               & \textbf{0.0321}               &     14.23\%                               \\
		& R@20                    & 0.0355               & 0.0632               & 0.0646               &  & 0.0662               & 0.0796               & 0.0671               & \underline{0.0901}                 & 0.0637               & 0.0870               & \textbf{0.0989}               &        9.77\%                             \\
		& N@20                    & 0.0144          & 0.0256        & 0.0263       &                      & 0.0266          & 0.0298         & 0.0272       & 0.0339                           & 0.0257             & \underline{0.0349}              & \textbf{0.0401}             &    14.89\%                                 \\ \midrule
		\multirow{4}{*}{Clothing}                       & R@10                    & 0.0119          & 0.0199        & 0.0225       &                      & 0.0220          & 0.0301         & 0.0205       & 0.0403                           & 0.0188             & \underline{0.0422}              & \textbf{0.0553}             &   31.04\%                                  \\
		& N@10                    & 0.0057          & 0.0098        & 0.0109       &                      & 0.0109          & 0.0127         & 0.0100       & 0.0172                             & 0.0091             & \underline{0.0201}              & \textbf{0.0256}             &       27.36\%                              \\
		& R@20                    & 0.0189     & 0.0308   & 0.0339  &                      & 0.0342     & 0.0450    & 0.0320  & 0.0612                       & 0.0301        & \underline{0.0639}         & \textbf{0.0808}        &   26.45\%                                  \\
		& N@20                    & 0.0075          & 0.0125        & 0.0137       &                      & 0.0139          & 0.0164      & 0.0128       & 0.0223                           & 0.0118             & \underline{0.0254}              & \textbf{0.0318}             &     25.19\%                                \\ 
		\bottomrule
		\end{tabular}}
		\caption{Performance comparisons of MDSRec and other baselines on five datasets. The best result is in boldface and the second best is underlined. Improvement is obtained between MDSRec and the best result in baselines.}
		\label{tab:performance}%
\end{table*}

\subsection{Experimental Setup}
\subsubsection{Datasets}
We conduct evaluation experiments on five publicly available benchmark datasets from widely-used Amazon platform\footnote{http://jmcauley.ucsd.edu/data/amazon/links.html}, which contains reviews from millions of Amazon customers. We collect (a) $\emph{Industrial and Scientific}$, (b) $\emph{Prime Pantry}$, (c) $\emph{Baby}$, (d) $\emph{Sports and Outdoors}$, and (e) $\emph{Clothing, Shoes and Jewelry}$ to train and evaluate our method. We refer to them separately as \textbf{Scientific}, \textbf{Pantry}, \textbf{Baby}, \textbf{Sports}, \textbf{Clothing} for brevity. Table~\ref{tab:dataset} summarizes the statistics results of these five datasets. Among them, the longest sequence lengths for datasets \textbf{Scientific} and \textbf{Pantry} are both $50$, while the longest sequences for \textbf{Baby}, \textbf{Sports}, and \textbf{Clothing} are $124$, $295$, $135$, respectively.
\subsubsection{Evaluation Protocols}
The performance of our MDSRec on the testing set is evaluated by two commonly used protocols: Recall (R@\textit{N}) and Normalized Discounted Cumulative Gain (N@\textit{N}). Recall@\textit{N} focuses on how many correct items are recommended, while NDCG@\textit{N} accounts for the ranking quality of correct items. We truncate the ranked list by setting $N$ at $\{10,20\}$. After training, the learned recommendation model can get a ranked top-\textit{N} list from all items to evaluate the two protocols. 

\subsubsection{\textbf{Baselines}}
We compare our MDSRec with the following competitive methods, divided into two groups: 
1) ID-based Sequential Recommendations: \textbf{GRU4Rec}~\cite{gru4rec}, \textbf{SASRec}~\cite{sasrec}, 
\textbf{BERT4Rec}~\cite{bert4rec}. 
2) Modality-based Sequential Recommendations:
\textbf{GRU4RecF},
\textbf{SASRecF}, 
\textbf{FDSA}~\cite{fdsa}, 
\textbf{UniSRec} ~\cite{unisrec}, 
\textbf{MMMLP}~\cite{mmmlp}, 
\textbf{MissRec}~\cite{missrec}. 
Here, \textbf{GRU4RecF} and \textbf{SASRecF} extend \textbf{GRU4Rec} and \textbf{SASRec} by combing multimodal features with ID embeddings, respectively.

\subsubsection{\textbf{Parameter Settings}}
Our model is implemented in Pytorch\footnote{https://pytorch.org}. For each user sequence in all datasets, we select the last item to construct test set and the one before it for the validation set. The remaining items are included in the training set.
For a fair comparison, we optimize all models via the Adam~\cite{adam} optimizer with the fixed embedding size $300$ and the mini-batch size $512$. Besides, we search the learning rate from $\{1e^{-4}, 1e^{-3}, \dots, 1e^{-1}\}$, the neighbor number $H$ in modal-aware relation graph from $\{0, 5, 10, 15, 20, 25, 30, 35, 40, 45, 50\}$, the center number $k$ from $\{2, 4, 8, 16, 32, 64, 128\}$. Unless otherwise stated, we set $\mu^{id}=16$, $\mu^m=0.2$, $k=32$, $H=10$. The early stop mechanism with a patience of $10$ is applied in our training process to alleviate overfitting problems. 

\subsection{Performance Comparison}
Table~\ref{tab:performance} reports the performance comparisons of MDSRec and all baselines in terms of Recall and NDCG on five datasets. From the table, we have the following observations:
\begin{itemize}[leftmargin=1em] 
  \item MDSRec almost achieves significant improvements over all baselines across five datasets.  Specifically, the average improvement of MDSRec on Clothing dataset can reach about 25\%, compared to MissRec. The results demonstrate the superiority of MDSRec in modeling modality-related differences. 
  
  \item By introducing modal features, GRU4RecF and SASRecF outperform their counterpart (i.e., GRU4Rec and SASRec). Meanwhile, other modality-based SR methods (e.g., FDSA, MissRec) also achieve superior performance than ID-based methods via more efficient modeling of modal features. The results indicate the effectiveness of introducing modal features to modeling representations of users and items. 
  
  \item Compared with GRU4RecF, SASRecF using Transformer as the backbone network achieves better results. Similarly, MissRec performs better than MMMLP that using MLP as backbone network, which verify that the Transformer architecture is beneficial for sequence manner modeling. 
\end{itemize}

\begin{table}[t]
	\small
	\centering
	\setlength{\tabcolsep}{1.2mm}{
		\begin{tabular}{l|c|c|c|c|c|c}
			\toprule
			\textbf{Datasets} & \multicolumn{2}{c|}{\textbf{Scientific}} & \multicolumn{2}{c|}{\textbf{Pantry}} & \multicolumn{2}{c}{\textbf{Baby}} \\
			\midrule
			\textbf{Methods} & \textbf{R@20} & \textbf{N@20} & \textbf{R@20} & \textbf{N@20} & \textbf{R@20} & \textbf{N@20} \\
			\midrule
			\textbf{w/o DIS} & 0.1734  & 0.0824  & 0.1192  & 0.0467  & 0.0894  & 0.0350  \\
			\textbf{w/o CRE} & 0.1714  & 0.0838  & 0.1153  & 0.046  & 0.0892  & 0.0347  \\	
			\textbf{w/o MRGC} & 0.1510  & 0.0728  & 0.0912  & 0.0387  & 0.0761  & 0.0314  \\
			\textbf{w/o ICA} & 0.1630  & 0.0833  & 0.0972  & 0.0411  & 0.0817  & 0.0343  \\
			\midrule
			\textbf{MDSRec} & \textbf{0.1761} & \textbf{0.0873} & \textbf{0.1201} & \textbf{0.0484} & \textbf{0.0905} & \textbf{0.0364} \\
			\bottomrule
	\end{tabular}}%
	\caption{The effectiveness of different variants of MDSRec.}
	\label{tab:ablation}%
\end{table}%

\subsection{Ablation Studies}
Table~\ref{tab:ablation} shows the results of ablation studies of MDSRec on Scientific, Pantry and Baby datasets. Specifically, 
\textbf{w/o DIS} denotes that the relative position of items within the sequence is not considered when extracting co-occurrence relation; 
\textbf{w/o CRE} is a variants that constructs the modal-aware relation graph using original modal features without behavioral information.
\textbf{w/o MRGC} abandons item relation graph construction module and directly uses original modal features as input for user sequence representation learning.
\textbf{w/o ICA} removes interest-centralized attention mechanism. 
Note that we have also conducted the same experiments on Sports and Clothing, the results exhibit similar trend and hence are omitted here due to space concern. From Table~\ref{tab:ablation}, we can observe: 
\begin{itemize}[leftmargin=1em] 
	\item Both \textbf{w/o MRGC} and \textbf{w/o ICA} cause a significant performance decline of MDSRec, demonstrating the effectiveness of modeling differences in user preferences and item relation across modalities. Moreover, the performance of \textbf{w/o MRGC} is worse than \textbf{w/o ICA}, which indicates that capturing item differentiated semantic relationships is more meaningful for boosting performance.
	
	\item The performance of \textbf{w/o CRE} decreases by 2.67\%, 4.00\%, and 1.44\% in term of R@$20$ respectively on the three datasets compared to MDSRec. This result verifies the necessity of introducing behavioral signals to construct modal-aware relation graph.
	
	\item The performance decreases of $\textbf{w/o DIS}$ indicate that the relative position of items in the sequence is very beneficial for accurately measuring their co-occurrence relationship. 
\end{itemize}

\begin{figure}[t]
	\centering
	\begin{minipage}[t]{0.5\textwidth}
		\centering
		\includegraphics[width=\textwidth,height=0.35\textwidth]{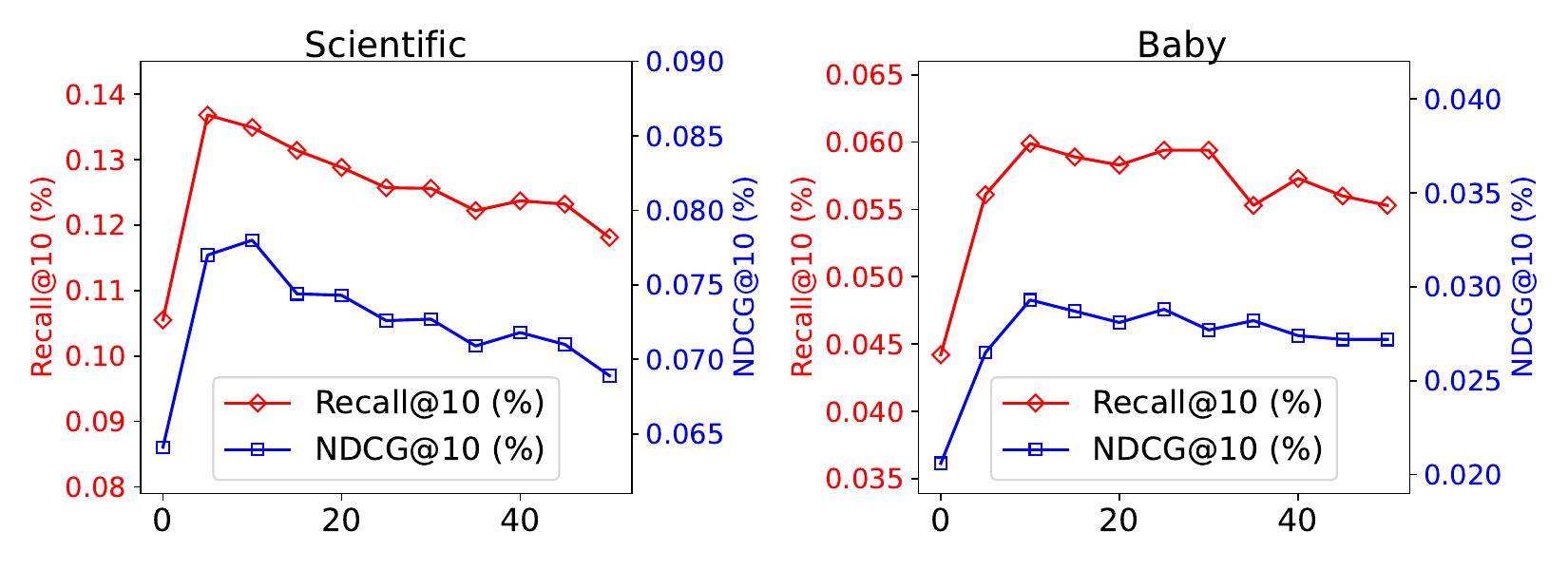}
	\end{minipage}
	\caption{The impact of different neighbor number $H$.}
	\label{fig:H}
\end{figure}

\begin{figure}[t]
	\centering
	\begin{minipage}[t]{0.5\textwidth}
		\centering
		\includegraphics[width=\textwidth,height=0.35\textwidth]{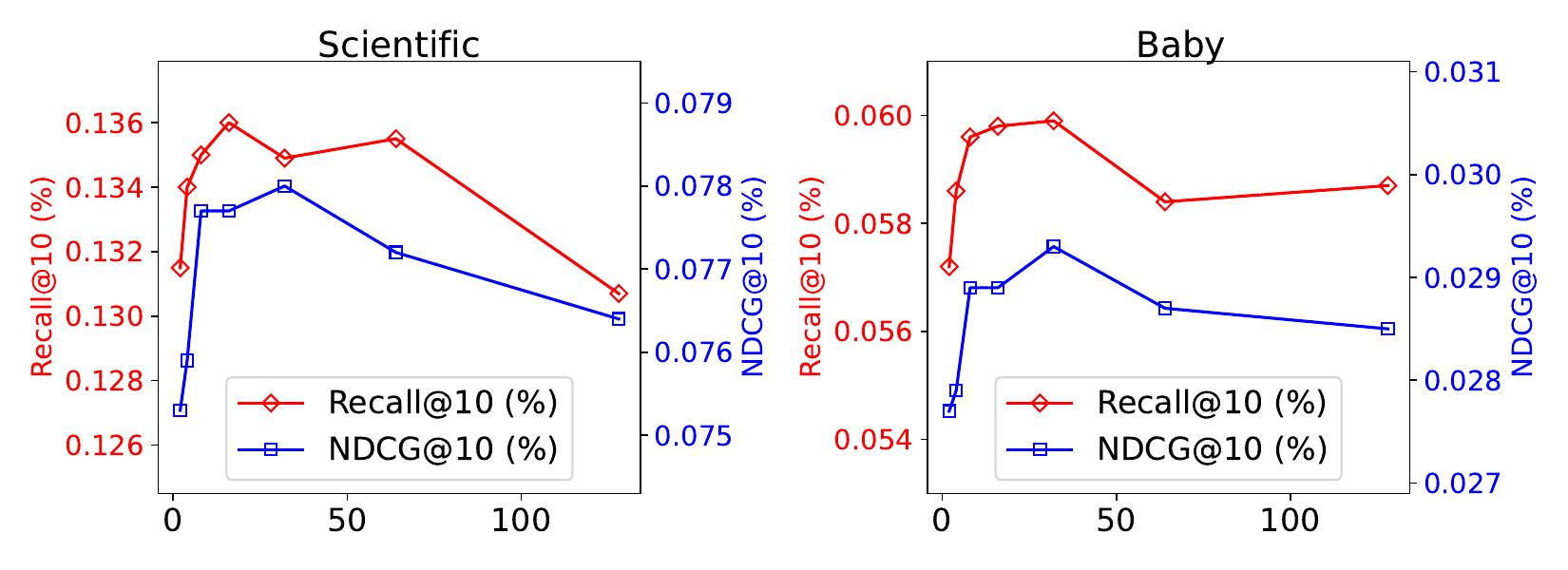}
	\end{minipage}
	\caption{The impact of different center number $k$.}
	\label{fig:k}
	\vspace{-1em}
\end{figure}

\subsection{In-depth Analysis}
\subsubsection{Impact of the neighbor number $H$}
To explore the impact of the neighbor number $H$ on model performance, we record the results of MDSRec with different $H$ on Scientific and Baby datasets. From the results in Figure~\ref{fig:H}, we can observe that as the $H$ increases, the model performance initially improves to a optimal results and then gradually declines. The reason is that an appropriate number of neighbors can enrich the semantic representation of items, but too many neighbors may introduce irrelevant semantic noise, negatively impacting performance. In practice, we set $H=5,10,10,20,15$ on Scientific, Pantry, Baby, Sports and Clothing datasets for optimal results, respectively.

\subsubsection{\textbf{Impact of the center number $k$}}
To further verify the effect of center number $k$, we report the performance of MDSRec with various $k$ in Figure~\ref{fig:k}. From the results, we can observe that the performance is optimal when $k$ increases to the range of $16-32$. Continuing to increase the number $k$ of centers may cause user interests to become more diffuse, making it more challenging to accurately extract the user’s primary interests. In practice, we obtain the best performance by setting $k=16,16,32,32,64$ on Scientific, Pantry, Baby, Sports and Clothing datasets, respectively.

\begin{figure}[t]
	\centering
	\hspace*{-1.5em}
	\begin{minipage}[t]{0.52\textwidth}
		\centering
		\includegraphics[scale=0.65]{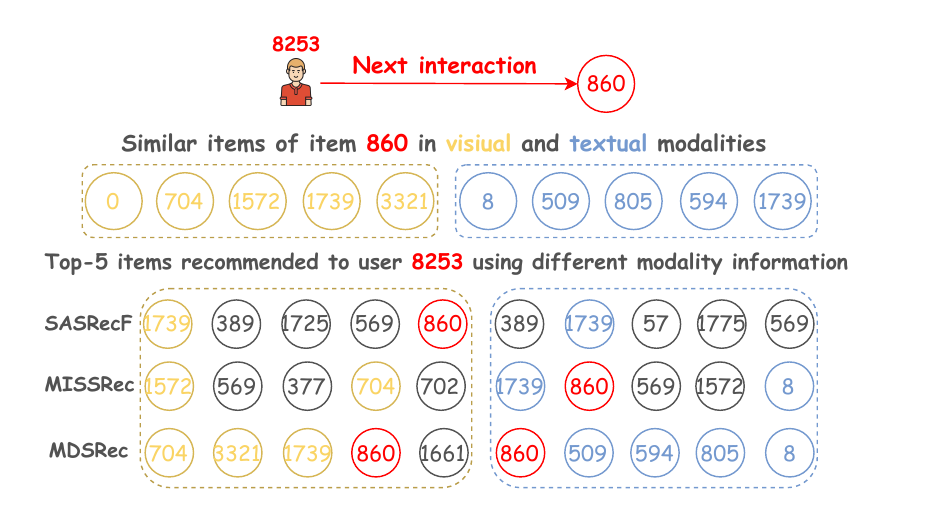}
	\end{minipage}
	\caption{Case studies of multimodal difference learning.}
	\label{fig:case}
\end{figure}

\subsection{Case studies}
To quantify the rationality of modality difference learning, we select a user $u_{8253}$ and its next interaction item $x_{860}$ from the Baby dataset, and then analyze the recommendation results of MDSRec and two baselines: SASRecF and MissRec. As shown in Figure~\ref{fig:case}, we identify five neighbors that are semantically similar to item $x_{860}$ in both visual and textual modalities. Obviously, the item $x_{1739}$ is the shared item, while other four neighbors differ across the two modalities. This indicates the significant relation differences for item $x_{860}$ between the modalities.
By analyzing the top-$5$ recommendations of SASRecF, MissRec and MDSRec on two modalities, we find that SASRecF and MissRec typically generate lower prediction rankings for the next interaction item $x_{860}$. However, the rankings of item $x_{860}$ from MDSRec is higher, i.e., fourth in visual modality and first in textual modality. Besides, the recommended items by SASRecF or MissRec under the visual and textual modalities are mostly overlapping, indicating that they fail to accurately capture the knowledge differences between modalities. Our proposed MDSRec yields differentiated recommendation results across different modalities, and the results include semantic neighbors of the items.
These results demonstrates that MDSRec can capture and leverage the differences in item relation and user interests across modalities to facilitate item recommendations.

\section{Conclusion} 
\label{sec:conclusion}

In this work, we proposed a new sequential recommendation method MDSRec, which captures and utilizes the differences in item relation and user interests across modalities to facilitate item recommendations. Specifically, we extracted item relation structures via behavior sequence and modal features to enhance item representations. Besides, we introduced a interest-centralized attention mechanism to mine user differentiated interests across modalities. 
Experiments on five real-world datasets demonstrate the superiority of MDSRec and the effectiveness of learning modality differences.
For future work, We plan to utilize the multimodal data of items to explore the generation of interpretable recommendation results.

\section{Acknowledgements} 
\label{secacknowledgements}
The constructive comments from the reviewers have been of great help to our work, and we are very grateful.

\bibliography{aa}

\clearpage

\end{document}